\documentclass[aps,prl,twocolumn,showpacs,superscriptaddress,nofootinbib,preprintnumbers]{revtex4-2}
\usepackage[bottom]{footmisc} 

\usepackage{graphicx}
\usepackage{subcaption}
\usepackage{amsmath}
\usepackage{mathrsfs}
\usepackage[dvipsnames]{xcolor}
\usepackage{hyperref}
\usepackage{cleveref}
\crefname{equation}{Eq.}{Eqs.}
\usepackage{comment}
\usepackage{gensymb}
\usepackage{style}
\usepackage{slashed}

\usepackage{dcolumn}
\usepackage{bm}
\usepackage{nccmath}

\hypersetup{colorlinks=true, linkcolor=blue, citecolor=blue, urlcolor=blue, filecolor=blue}

\newcommand{\bea}{\begin{equation}\begin{aligned}}
\newcommand{\eea}{\end{aligned}\end{equation}}

\begin{document}

\title{Cosmological Signals from Finite-Lifetime Domain Wall Networks}

\author{Fengwei Yang\,\orcidlink{0000-0001-9873-6259}}
\email{fyang@uni-mainz.de}
\affiliation{Department of Physics and Astronomy, University of Notre Dame, South Bend, IN 46556}
\affiliation{Johannes Gutenberg-Universit\"{a}t Mainz, 55099 Mainz, Germany}

\author{Kylar Greene\,\orcidlink{0000-0002-2711-7191}}
\email{kylar.cosmo@pm.me}
\affiliation{Department of Physics and Astronomy, Seoul National University, 1 Gwanak-ro, Gwanak-gu, Seoul 08826, Korea}

\author{Aurora Ireland\,\orcidlink{0000-0001-5393-0971}} 
\email{anireland@stanford.edu}
\affiliation{Leinweber Institute for Theoretical Physics, Stanford University, Stanford, CA 94305}


\author{Yuhsin Tsai\,\orcidlink{0000-0001-7847-225X}}
\email{ytsai3@nd.edu}
\affiliation{Department of Physics and Astronomy, University of Notre Dame, South Bend, IN 46556}

\begin{abstract}
We study cosmological perturbations sourced by metastable domain wall (DW) networks. Using the unconnected segment model, we derive analytic expressions for the scalar and tensor power spectra in the long-wavelength regime, finding a fixed ratio between their amplitudes during the scaling regime. These spectra grow as the network evolves, driven by the increasing correlation length and slower redshifting of the wall energy density relative to radiation, and peak at a scale set by the network's collapse time. Comparing the scalar spectrum with observational constraints, we derive bounds on the DW energy density at collapse, which we apply to $\mathbb{Z}_2$ and axion models. When these bounds are saturated, the resulting CMB $B$-mode signal can be comparable to or larger than the inflationary signal for a tensor-to-scalar ratio $r \sim 0.03$, potentially bringing it within reach of future CMB polarization experiments. 
\end{abstract}

\maketitle

\section{Introduction}

Domain walls (DWs) are topological defects that can form from the spontaneous breaking of discrete symmetries in the early universe~\cite{Zeldovich:1974uw,Kibble:1976sj,Vilenkin:2000jqa}. If the discrete symmetry is exact, the DW network is stable; if it is only approximate, the network is metastable, with a lifetime set by the hierarchy between the explicit and spontaneous breaking scales. DWs that survive sufficiently long can leave observable signatures in the cosmic microwave background (CMB)~\cite{Lazanu:2015fua,Sousa:2015cqa}, the stochastic gravitational wave background~\cite{Saikawa:2017hiv,Gruber:2024pqh}, and large-scale structure~\cite{Lola:1993qu,Winckler:2025hbc}.

Because DW energy density redshifts as $\rho_{\rm DW} \propto a^{-2}$ during radiation domination, whereas $\rho_R \propto a^{-4}$, their fractional contribution to the energy budget grows as $f_{\rm DW} \equiv \rho_{\rm DW}/\rho_{R} \propto a^2$. Stable DW networks thus face strong cosmological constraints. Consider, for example, a network formed during a symmetry-breaking phase transition at a temperature $T_i$ in the radiation dominated era. The initial energy fraction is
\bea
    f_{\rm DW}^i = \frac{\sigma H_i}{\xi \rho_R^i} \sim \frac{T_i}{\xi M_{\rm Pl}} \,,
\eea
where $H_i$ is the Hubble rate at formation and $\xi \sim 1$ parametrizes the correlation length growth rate. The wall tension $\sigma$ is set by the symmetry-breaking scale, here assumed to be of order the transition temperature, $\sigma \sim T_i^3$. Requiring that the energy density in DWs remains sub-dominant $f_{\rm DW} \lesssim 0.1$ through recombination gives an approximate upper bound on the formation temperature,
\bea
   T_{i}^\text{max} \sim (0.1 M_{\rm Pl} T_{\rm rec}^2)^{1/3}\sim 100\,\mathrm{MeV} \,. 
\eea
This illustrates the strong constraints on stable DW networks and motivates us to consider \textit{unstable} DWs, which can collapse before dominating the universe, allowing for higher spontaneous symmetry-breaking scales and potentially observable signatures on scales probed by both the CMB and small-scale observations.

Here we study the perturbations generated by finite-lifetime DWs both before and at collapse. To describe the cosmological evolution of the network, we use the unconnected segment model (USM), originally proposed for cosmic strings in~\cite{Albrecht:1997nt,Albrecht:1997mz,Pogosian:1999np} and generalized to DW networks in~\cite{Sousa:2015cqa}. The inhomogeneous distribution of the walls, together with the intrinsic anisotropy of their geometry, sources scalar and tensor perturbations through the DW stress-energy tensor. 

Using this framework, we compute the scalar $\mathcal{P}_\zeta^{\rm DW}$ and tensor $\mathcal{P}_h^{\rm DW}$ power spectra, obtaining for the first time analytical expressions in the long-wavelength regime. 
By comparing with existing constraints on curvature perturbations, we derive upper bounds on the allowed fractional energy density in DWs at the time of collapse, $f^c_{\rm DW}$. We illustrate how these bounds can be translated into constraints on specific symmetry-breaking scenarios, such as DW production from spontaneous $\mathbb{Z}_2$ breaking or from axion models with instanton-induced oscillatory potentials. 

The bound on $f^c_{\rm DW}$ also determines the maximum allowed gravitational wave (GW) signal imprinted on the CMB $B$-mode spectrum. We find that the resulting $B$-mode signal sourced by DW perturbations can be comparable to, or even exceed, that from inflation with a tensor-to-scalar ratio of $r \sim 0.03$, placing it within reach of future $B$-mode observations.

\section{Scaling Domain Wall Network }

We describe the large-scale evolution of the DW network using the velocity-dependent one-scale (VOS) model~\cite{Avelino:2005kn,Martins:2016ois}. In this framework, the network is characterized by two parameters: the comoving correlation length $\ell_{\rm corr}$ and the root-mean-square wall velocity $v$. In the scaling regime, the competition between Hubble friction, wall tension, and energy loss drives the network towards a self-similar scaling solution. The velocity approaches a constant, and the correlation length grows linearly with conformal time $\eta$ as,
\bea\label{eq:xi:def:implicit}
    \ell_{\rm corr} = \xi \eta \,,
\eea
with $\xi$ and $v$ determined by the properties of the network.

To calculate the perturbations sourced by the DW network, we use the USM, which approximates the network as an ensemble of uncorrelated flat, square wall segments of negligible thickness. Each segment has comoving length $\ell_{\rm corr}$ and moves with a velocity $v$ normal to its surface. The stress-energy tensor of an individual segment reads
\bea\label{eq:Tmunu}
    T^{\mu\nu} \!=\!- \frac{\sigma}{\sqrt{-g}} \!\int\!\! d^3\zeta \, \delta^{(4)}\! \big( x^\mu - x^\mu_{\rm DW} (\zeta^a) \big) \left( \sqrt{-h} h^{ab} x_{,a}^\mu x_{,b}^\nu \right) \,,
\eea
where $g_{\mu\nu} = a^2\text{diag} (-1, 1, 1, 1)$ and $g=\det(g_{\mu\nu})$. The coordinates $\zeta^a$ parameterize the wall worldsheet, with $\zeta^0=\eta$ and $\zeta^{1},\zeta^2$ spatial. The induced metric on the worldsheet is $h_{ab} = g_{\mu\nu} x_{,a}^\mu x_{,b}^\nu$, with $h = \det(h_{ab})$. Finally, the delta function localizes the stress-energy on the DW trajectory, $x^\mu_{\rm DW}(\zeta^a) = (\eta, \textbf{x}_{\rm DW}(\zeta^a))$.

\section{Power Spectra}

Using the stress-energy tensor in Eq.~(\ref{eq:Tmunu}) and taking the ensemble average over segment positions, orientations, and velocity directions, we obtain the unequal-time correlators for the Fourier modes of $T^{00}$ and $T^{ij}$. We then use these to compute the scalar and tensor power spectra sourced by the DW network at radiation domination; see the Supplemental Material for details. Presuming $\delta\rho_{\rm DW} \ll \bar{\rho}_{\rm tot}$, the curvature power spectrum in the long-wavelength regime $\xi k \eta \ll 1$ reads
\bea\label{eq:Pzeta_dimless}
    \mathcal{P}^{\rm DW}_\zeta(\eta,k)=\frac{\gamma^2(f^{i}_{\rm DW})^2}{32\pi^2}(\xi k \eta_{i})^3 \left(\frac{g_{\star}}{g_{\star,i}}\right)^{-\frac{1}{3}} \left(\frac{\eta}{\eta_{i}}\right)^7 \,,
\eea
where the dimensionless power spectrum for a Fourier mode $X_{\mathbf{k}}$ is defined via
\bea
    \mathcal{P}_X(\eta,k) (2\pi)^3 \delta^{(3)}(\mathbf{k}-\mathbf{k}') = \frac{k^3}{2\pi^2} \langle X_{\mathbf{k}}(\eta) X^*_{\mathbf{k}'}(\eta) \rangle \,,
\eea
$\gamma=(1-v^2)^{-1/2}$ is the Lorentz factor, $g_\star(T)$ is the effective number of relativistic degrees of freedom, and $g_{\star,i} \equiv g_\star(T_i)$. For simplicity, we approximate $g_\star(T) \simeq g_{\star,s}(T)$ throughout. The enhancement $(\eta/\eta_{i})^7$ reflects the growth of the comoving correlation length together with the slower redshifting of the DW energy density.

Meanwhile, the tensor power spectrum in the long-wavelength regime while the DW network is still present ($\eta<\eta_c$) reads
\bea\label{eq:Ph_dimless}
   \mathcal{P}^{\rm DW}_h(\eta,k) = 0.386\,\mathcal{P}^{\rm DW}_\zeta(\eta,k) \,,
\eea
where $\mathcal{P}^{\rm DW}_h$ includes the sum over both polarizations. Notice that the scalar and tensor spectra are directly proportional in this limit, with an $\mathcal{O}(1)$ coefficient of proportionality. This contrasts with inflation, where the tensor power spectrum is typically suppressed relative to the scalar by a small value for $r$. The comparable amplitudes here reflect that both scalar and tensor perturbations are sourced by the same inhomogeneous and anisotropic DW stress-energy tensor.

To better understand the scaling behavior, we use Eq.~(\ref{eq:xi:def:implicit}) to re-write the conformal time and wavenumber dependence as
\bea\label{eq:klcscale}
    \mathcal{P}^{\rm DW}_{\zeta,h}(\eta,k) \propto \big( k \ell_{\rm corr}(\eta) \big)^3 \left( \frac{\eta}{\eta_i} \right)^4 \,.
\eea
The factor $\left( k\,\ell_{\rm corr}(\eta)\right)^3$ gives the usual causality-limited scaling~\cite{Greene:2026one,Elor:2023xbz}, with power suppressed on length  scales larger than the network coherence length, $k \ll \ell_{\rm corr}^{-1}$. The factor $(\eta/\eta_{i})^4$ comes from the relative redshift of the squared DW energy density compared to radiation.

We denote the network collapse time by $\eta_c$, and presume that after collapse the energy is converted into gravitationally coupled dark radiation; stronger constraints can arise for decays to Standard Model radiation~\cite{Elahi:2026ldn}. For modes that enter the horizon before $\eta_c$ (i.e., $k^{-1}<\eta_c$), we evaluate the power spectra in Eqs.~(\ref{eq:Pzeta_dimless}) and (\ref{eq:Ph_dimless}) at horizon re-entry, $\eta = k^{-1}$. For modes that remain super-horizon at collapse ($k^{-1}>\eta_c$), we instead evaluate the scalar spectrum at $\eta=\eta_c$, since afterwards the scalar perturbations are effectively frozen. For tensor modes, by contrast, the power spectrum at collapse does not directly give the final frozen spectrum. Once the DW source disappears, super-horizon tensor modes evolve as a linear combination of ``constant'' and decaying modes. The amplitude of the constant mode depends on both the tensor perturbation and its time derivative at collapse, leading to an additional enhancement in the final super-horizon tensor spectrum~\cite{Hook:2020phx}.

In the scalar case, the super-horizon power spectrum thus reads
\bea\label{eq:Pzeta_re}
    \mathcal{P}_\zeta(k) \simeq \mathcal{N} \begin{cases} \left(\frac{g_{\star,i}}{g_{\star}}\right)^{\frac{1}{3}} (k\eta_{i})^{-4} & \eta_c^{-1} \! \le \! k \! \leq \! (\xi\eta_{i})^{-1} \,, \\ \left(\frac{g_{\star,i}}{g_{\star,c}}\right)^{\frac{1}{3}} \!(k\eta_{i})^3 \! \left(\frac{\eta_c}{\eta_{i}}\right)^7 & k \!<\! \eta_c^{-1} \,, \end{cases} 
\eea
where we have defined the normalization
\bea\label{eq:normalization}
    \mathcal{N} = 10^{-13}\, \gamma^2 \! \left(\frac{\xi}{0.63}\right)^3 \! \left(\frac{f^i_{\rm DW}}{10^{-5}}\right)^2 \,.
\eea
We see that the spectrum can be well-approximated by a broken power law, peaking at $k_p = \eta_c^{-1}$. For $k < \eta_c^{-1}$, the spectrum follows the usual causality-limited scaling $\propto k^3$. For $k > \eta_c^{-1}$, modes enter the horizon before the network collapses, when the DW energy density fraction is smaller, suppressing the resulting perturbations. 

In the tensor case, the corresponding constant-mode spectrum is
\bea\label{eq:Ph:constant}
    & \mathcal{P}_{h,\text{const}}(k) \simeq 0.386 \mathcal{N} \\
    & \times \begin{cases} \left(\frac{g_{\star,i}}{g_{\star}}\right)^{\frac{1}{3}} (k\eta_{i})^{-4} & \eta_c^{-1} \! \le \! k \! \leq \! (\xi\eta_{i})^{-1} \,, \\ \mathcal{E}(k \eta_c) \left(\frac{g_{\star,i}}{g_{\star,c}}\right)^{\frac{1}{3}} \!(k\eta_{i})^3 \! \left(\frac{\eta_c}{\eta_{i}}\right)^7 & k \!<\! \eta_c^{-1} \,. \end{cases} 
\eea
Comparing with the scalar spectrum, we note that in addition to the overall proportionality factor $0.386$ already present in the scaling regime, there is an additional enhancement factor $\mathcal{E}(k\eta_c)$ for the super-horizon tail. This new factor accounts for the additional evolution of super-horizon tensor modes after the DW source disappears. Its explicit form can be found in the Supplemental Materials. We simply note here that it is weakly dependent on $k$, varying between $\mathcal{E}(0) \simeq 22.7$ in the asymptotic tail ($k\eta_c \rightarrow 0$) and $\mathcal{E}(1) \simeq 10.8$ near the peak ($k\eta_c \sim 1$). Since the derivation required taking the long-wavelength approximation in the UETC, however, the enhancement factor at the peak should be regarded as an estimate. Thus, while scalar and tensor spectra are directly proportional while the network is scaling, the subsequent evolution of tensor modes modifies their relative amplitudes in the final frozen spectra.

Note that our prescription neglects subsequent sub-horizon evolution of the DW network, which may further enhance perturbations at $k \gg \eta_c^{-1}$, e.g., through DW fragmentation. Modeling these effects requires numerical simulations beyond the scope of this work (see e.g.~\cite{Cyr:2025nzf}.) This should not affect the spectrum near the peak at $k=\eta_c^{-1}$ or at smaller $k$, which remain super-horizon at the time of decay. Our DW constraints, which come mainly from perturbations around the $k=\eta_c^{-1}$ peak, are therefore expected to be robust to the inclusion of such effects.

\section{Curvature Perturbation Constraints}

\begin{figure}[t!]
    \centering
    \includegraphics[width=1.0\linewidth]{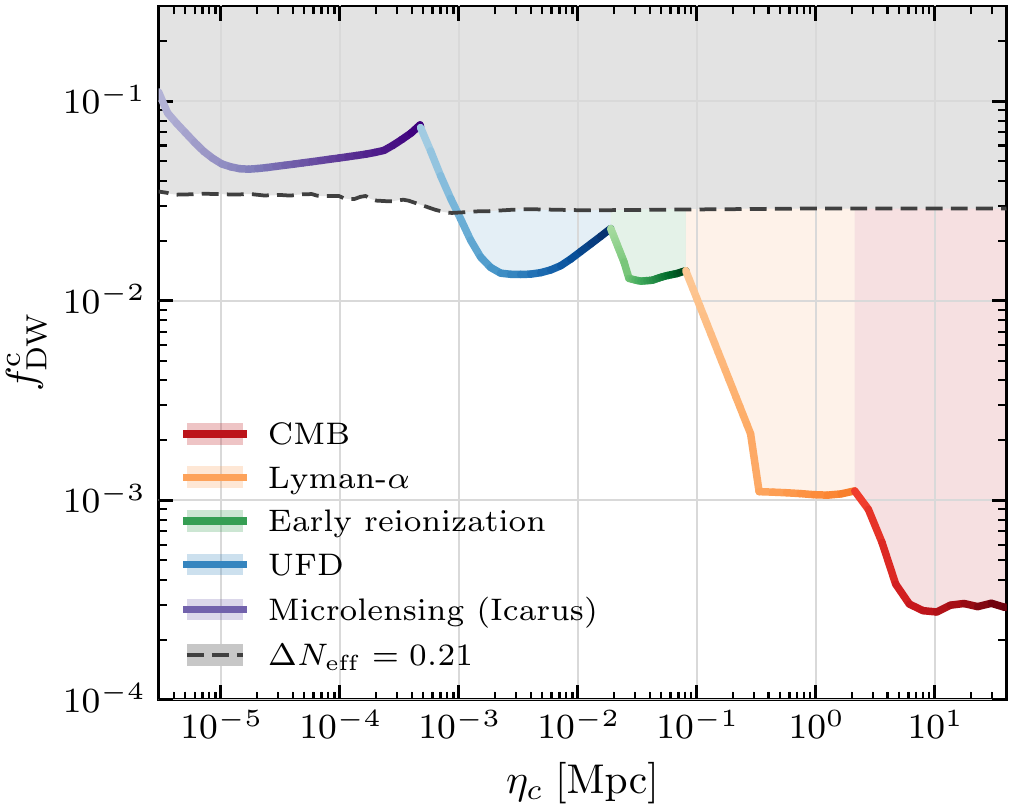}
    \caption{The $2\sigma$ upper bounds on the DW energy density fraction before collapse, $f_{\rm DW}^c$, as a function of the collapse time $\eta_c$, from CMB~\cite{Planck:2018jri} and Lyman-$\alpha$~\cite{2011MNRAS.413.1717B} constraints on the primordial curvature perturbation. We also show bounds from early reionization~\cite{Qin:2025ymc}, ultra-faint dwarf (UFD) heating~\cite{Graham:2024hah}, and microlensing of the Icarus star~\cite{Bringmann:2025cht}. We take $\xi=0.63$, with colors indicating the different constraints. The dashed curve shows the $2\sigma$ bound from $\Delta N_{\rm eff}$ (Planck 2018 TTTEEE+lowE+lensing)~\cite{Planck:2018vyg}. We restrict the plot to $\eta_c \leq 40 \, \text{Mpc}$ to ensure that collapse occurs during radiation domination, as assumed in our derivation.}
    \label{fig:vaOfetac_P_zeta}
\end{figure}

Given the analytic expression for $\mathcal{P}_\zeta^{\rm DW}(k)$ in Eq.~(\ref{eq:Pzeta_re}), one can constrain unstable DW networks by comparing their induced perturbations with various observational limits on primordial curvature perturbations. In particular, we constrain the fractional energy density in the DW network immediately before collapse, $f_{\rm DW}^c$, which is related to the initial fractional energy density as
\bea
    f^c_{\rm DW} = f^{i}_{\rm DW} \left(\frac{g_{\star,i}}{g_{\star,c}} \right)^{1/6} \! \left(\frac{\eta_c}{\eta_{i}}\right)^2 \,.
\eea
This fraction determines the peak amplitude of the scalar power spectrum evaluated at $k = \eta_c^{-1}$,
\bea
    \mathcal{P}_{\zeta}^{\rm DW} \left(k=\eta_c^{-1} \right) \simeq 10^{-5} \gamma^2 \left(\frac{\xi}{0.63} \right)^3 \left( \frac{f_{\rm DW}^c}{0.1} \right)^2 \,.
\eea
We obtain the upper bounds shown in Fig.~\ref{fig:vaOfetac_P_zeta} by comparing the total power spectrum $\mathcal{P}_{\zeta}^{\rm DW}(k)+\mathcal{P}_{\zeta}^{\rm ad}(k)$, using the standard $A_s=2.1\times10^{-9}$ at $k_*=0.05\,\mathrm{Mpc}^{-1}$ and $n_s=0.966$ for the adiabatic contribution~\cite{Planck:2018jri}, with various observational constraints on primordial curvature perturbations. The bounds based on the DW perturbations can be much stronger than the existing $\Delta N_{\rm eff}$ constraint for DWs decaying after $10^{-3} \,\text{Mpc}$.

These model-independent bounds can be translated into constraints on specific DW scenarios by relating the initial fractional energy density $f_{\rm DW}^i$ and collapse time $\eta_c$ to the model parameters. We illustrate this procedure for two representative case studies below. Further detail can be found in the Supplementary Materials.\\

\begin{figure}[t!]
        \centering
        \includegraphics[width=0.9\linewidth]{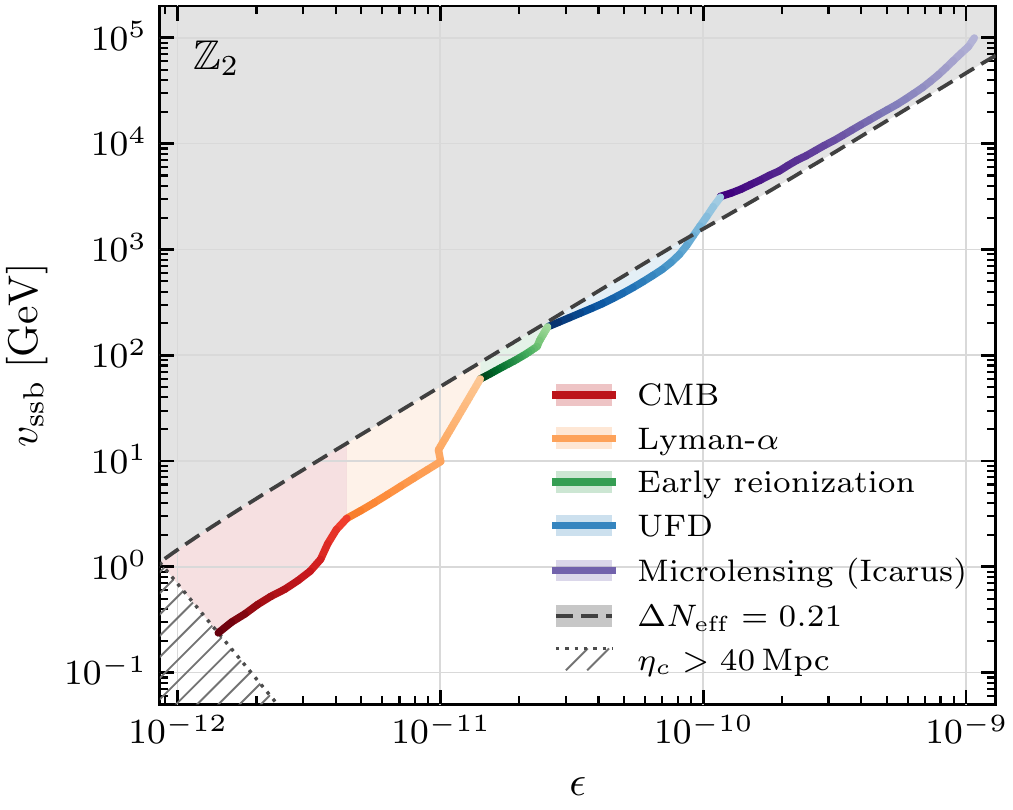}
    \\\vspace{0.2em}
    \includegraphics[width=0.95\linewidth]{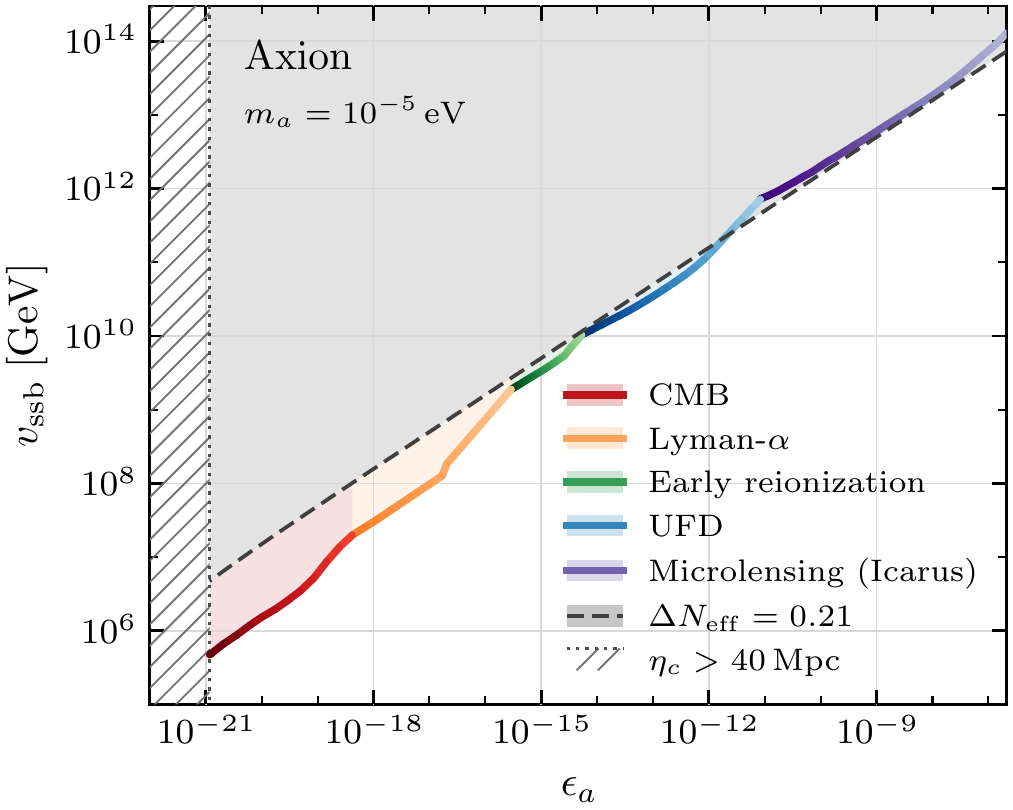}
    \caption{\emph{Upper:} bounds on unstable DWs in the $\mathbb{Z}_2$ model, with $v_{\rm ssb}$ the spontaneous symmetry-breaking scale and $\epsilon$ the ratio of the explicit symmetry-breaking scale to $v_{\rm ssb}$. We take $\lambda=1$, $\xi=0.63$, and wall velocity $v=0.5$. \emph{Lower:} bounds on unstable axion DWs, with the axion mass $m_a=10^{-5}\,\mathrm{eV}$, and $\epsilon_a$ parametrizing the explicit symmetry breaking. We take $N=2$, $\xi=0.63$, and $v=0.5$. We exclude regions with $\eta_c>40 \, \text{Mpc}$ (dotted) since our derivation assumes radiation domination. } 
    \label{fig:model}
\end{figure}

\noindent \textit{$\mathbb{Z}_2$ Model ---} Consider a real scalar field with a double-well potential described by the Lagrangian 
\bea
    \mathcal{L}=\frac{1}{2}(\partial_\mu\phi)^2-\frac{1}{4}\lambda\left(\phi^2-\frac{v_{\rm ssb}^2}{2}\right)^2 \,,
\eea
where $\lambda$ is the scalar self-coupling and $v_{\rm ssb}$ is the spontaneous symmetry-breaking scale. We will eventually add a small explicit breaking term $V_{\slashed{\rm sym}}$ to allow for DW collapse. For weak explicit breaking, the static wall profile is to leading order~\cite{Vilenkin:2000jqa}
\bea
    \phi(z) = \frac{v_{\rm ssb}}{\sqrt{2}} \tanh\left(\frac{\sqrt{\lambda}}{2}v_{\rm ssb} z\right) \,,
\eea
where $z$ is the coordinate normal to the wall. The corresponding wall tension is
\bea\label{eq.sigma}
    \sigma_{\mathbb{Z}_2} = \int_{-\infty}^{\infty} \!dz \,T_0^0(z)= \frac{\sqrt{\lambda}}{3} v_{\rm ssb}^3 \,.
\eea
For a phase transition occurring during radiation domination at $T_{i} \simeq \sqrt{3}\,v_{\rm ssb}$, we thus obtain 
\bea\label{eq:z2fDW}
    f^{i,\mathbb{Z}_2}_{\rm DW} \simeq   7.3\times10^{-21}\sqrt{\lambda}\left(\frac{0.63}{\xi}\right)\left(\frac{100}{g_{\star,i}}\right)^{\frac{1}{2}}\left(\frac{v_{\rm ssb}}{\rm GeV}\right) \,.
\eea

The explicit breaking term, which we parameterize as $V_{\slashed{\rm sym}}=\epsilon^3 v_{\rm ssb}^3\phi$ with $\epsilon \ll 1$, lifts the degeneracy between the two vacua, leading to a bias potential $V_{\rm bias} \equiv | V_{\slashed{\rm sym}}(\tfrac{v_{\rm ssb}}{\sqrt{2}})-V_{\slashed{\rm sym}}(-\tfrac{v_{\rm ssb}}{\sqrt{2}})| = \sqrt{2} \epsilon^3 v_{\rm ssb}^4$. DW network collapse occurs at a temperature $T_c$ determined by the condition that the wall pressure $|p_{\rm DW}|\sim \sigma H/\xi$ balances this bias energy density, $|p_{\rm DW}(T_c)| \simeq V_{\rm bias}$. The result is generically~\cite{Vilenkin:1981zs,Gelmini:1988sf}
\bea\label{eq:eta_c}
    \eta_c \simeq \sqrt{\frac{\sigma}{\beta(T_c) \xi \, V_{\rm bias}} } \,,
\eea
with $\beta(T) = a_0^2 H_0 \sqrt{\Omega_r}({g_\star}/{g_{\star,0}})^{-1/6}$, where the subscript ``$0$'' denotes quantities today and $\Omega_r$ is the radiation density parameter. We take the scaling parameter in this model to be $\xi = 0.63$, consistent with simulation~\cite{Leite:2011sc}. Using Eqs.~(\ref{eq:z2fDW}) and (\ref{eq:eta_c}), we can thus translate the bound in Fig.~\ref{fig:vaOfetac_P_zeta} into a constraint on $v_{\rm ssb}(\epsilon)$, as shown in the top panel of Fig.~\ref{fig:model}.\\

\noindent \textit{Axion DW ---} In axion models, instanton effects generate an oscillatory potential for the complex scalar $\Phi=\phi\,e^{i\theta}$, breaking the $U(1)$ symmetry down to $\mathbb{Z}_N$. The full Lagrangian is
\bea
    \mathcal{L}=|\partial_\mu\Phi|^2 \! -\frac{1}{4}\lambda\! \left(|\Phi|^2-\frac{v_{\rm ssb}^2}{2}\right)^2 \!\! -\frac{m_a^2 v_{\rm ssb}^2}{N^2}\left(1-\cos(N\theta)\right) \,,
\eea
and in particular, at low energies below the mass of the radial mode, the effective Lagrangian for the angular mode is
\bea
    \mathcal{L}_\theta=\frac{1}{2}v_{\rm ssb}^2(\partial_\mu\theta)^2-\frac{m_a^2v_{\rm ssb}^2}{N^2}(1-\cos(N\theta)) \,.
\eea
The spontaneous breaking of the discrete $\mathbb{Z}_N$ symmetry leads to DWs in the axion field~\cite{Sikivie:1982qv}, with profile
\bea
    \theta(z) = \frac{4}{N} \tan^{-1}\left(e^{m_a z}\right) \,,
\eea
where $m_a$ is the axion mass and $N$ is the DW number. Following a similar calculation as before, the wall tension is~\cite{Vilenkin:2000jqa} 
\bea
    \sigma_{a} = \frac{8m_a v_{\rm ssb}^2}{N^{2}}\,,
\eea
and the initial fractional energy density is
\bea
    f_{\rm DW}^{i,a} = \frac{64\pi v_{\rm ssb}^2}{\xi N^2 M_{\rm Pl}^2} =5.4 \times 10^{-19}N^{-2}\left(\frac{0.63}{\xi}\right)\left(\frac{v_{\rm ssb}}{10^9{\rm GeV}}\right)^2\,,
\eea
where we have estimated the DW network formation time $T_i$ by the onset of axion dynamics, i.e. $m_a \simeq 3 H(T_{i})$, leading to $T_{i} \simeq 15.66~{\rm TeV}(m_a/{\rm eV})^{1/2}(100/g_\star)^{1/4}$.

Since axion DWs are bounded by axion strings, their dynamics are more complicated than in the $\mathbb{Z}_2$ model. Nevertheless, lattice simulations find that the network can reach a scaling regime before collapse during radiation domination for $N=1$, and for an extended period for $N>1$~\cite{Hiramatsu:2012gg,Hiramatsu:2012sc}. For $N=1$, the scaling parameter is $\xi \simeq 1.04$; for $N=2$, $\xi \simeq 0.63$ (identical to the $\mathbb{Z}_2$ case); finally for larger $2 \leq N \leq 6$, $\xi_N/\xi_{N=2} \simeq 2/N$.

Adding an explicit breaking $V_{\slashed{\rm sym}} = \epsilon_a m_a^2 v_{\rm ssb}^2\cos(\frac{1}{2}N\theta)/N^2$, $N=2$ to the potential gives a bias $V_{\rm bias} = \epsilon_a m_a^2 v_{\rm ssb}^2/2$ between the $\theta=0$ and $\pi$ vacua. Using these parameters in the collapse time $\eta_c$ of Eq.~\eqref{eq:eta_c}, we can translate the constraint of Fig.~\ref{fig:vaOfetac_P_zeta} into that shown in the lower panel of Fig.~\ref{fig:model}.

\section{CMB $B$-Mode Signals}

\begin{figure}[t!]
    \centering
    \includegraphics[width=1.0\linewidth]{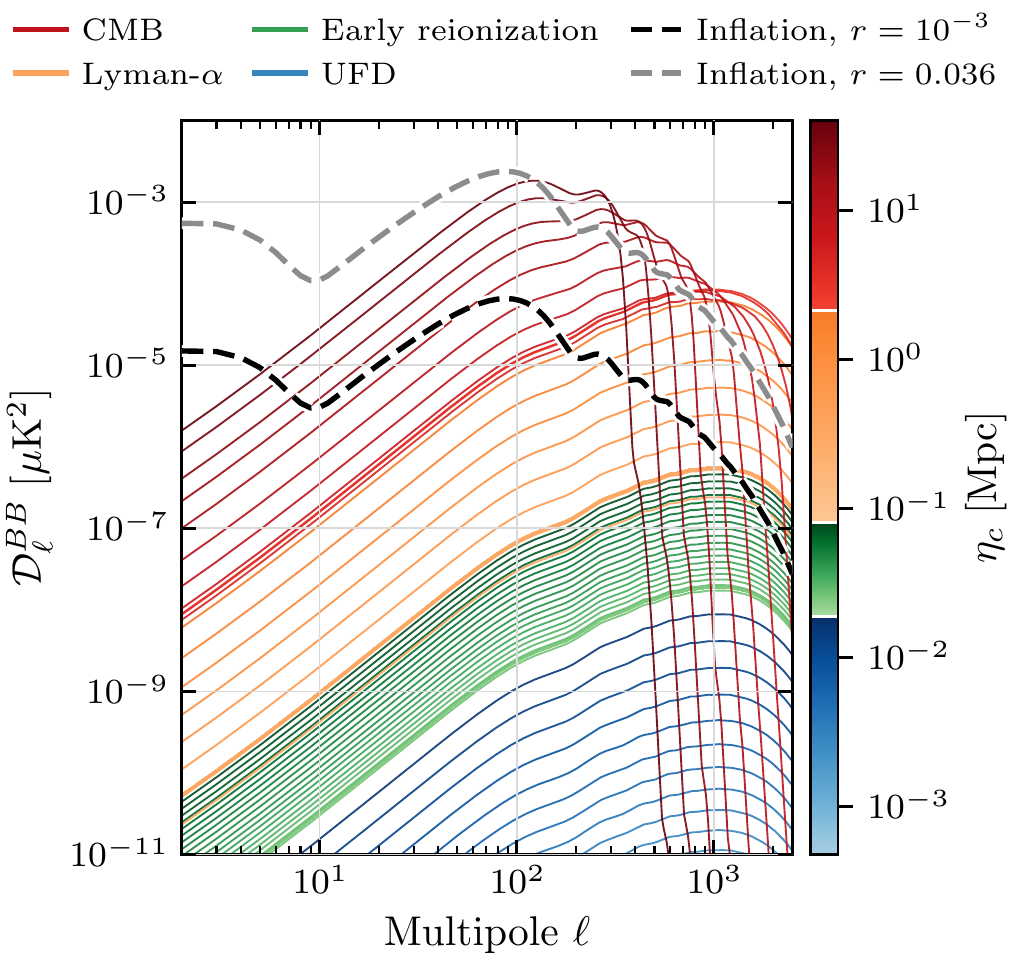}
    \caption{DW-sourced CMB $B$-mode spectra compared with scale-invariant inflationary predictions for $r=0.036$ (gray dashed) and $10^{-3}$ (black dashed). We take $\xi=0.63$, $v=0.5$, and fix $f_{\rm DW}^c$ to saturate the curvature-perturbation bound.
} \label{fig:B-mode_DW_r_0.001}
\end{figure}

As shown in Eq.~(\ref{eq:Ph_dimless}), DWs source scalar and tensor perturbations with a fixed ratio of power spectrum amplitudes on large scales. The bounds on the scalar power derived in the previous section thus also constrain the allowed tensor power, and by extension, the associated CMB $B$-mode signal. DWs can also generate a stochastic GW background (SGWB) through their evolution and collapse, which has been studied in various works~\cite{Saikawa:2017hiv,Gruber:2024pqh,Notari:2025kqq}. These SGWB constraints are complementary to those derived here, as our analysis focuses on the large-scale tensor perturbations that source CMB polarization.

In principle, the $B$-mode signal receives contributions from both tensor and vector perturbations sourced by the DW network~\cite{Sousa:2015cqa}. Since the modes relevant for the CMB enter the horizon after DW collapse ($k<\eta_c^{-1}$), we neglect the vector contribution, which decays after the source disappears. Additionally, we approximate the tensor perturbation by evolving the amplitude at horizon re-entry $h_\star(\bf{k})$ with the tensor transfer function $\mathcal{T}(\eta,k)$, i.e. $h(\eta,{\bf k}) \simeq h_\star({\bf k}) \mathcal{T}(\eta,k)$. This approximation is conservative in that it neglects the residual contribution from continued sourcing after horizon re-entry, which can slightly enhance the tensor amplitude and resulting $B$-mode signal.

The tensor power spectrum at horizon re-entry, $\mathcal{P}_h^{\rm DW}(k)$, obtained in Eq.~(\ref{eq:Ph:constant}), then enters the angular spectrum of $B$-mode polarization as 
\bea\label{eq:DlBB}
    \mathcal{D}_\ell^{\rm BB} = 18 \, \ell (\ell+1) \, T_0^2 \int_0^\infty  \frac{dk}{k} \mathcal{P}_h^{\rm DW}(k) \mathcal{F}_\ell(k)^2 \,,
\eea
where 
\begin{equation}\label{eq:Fl}
\begin{split}
    \mathcal{F}_\ell(k) = \int_0^{\eta_0} \!\!\! & d\eta \, V(\eta_0,\eta) \mathcal{S}_\ell(\eta,k) \! \int_0^{\eta} \!\!\! d\eta_1 \, V(\eta,\eta_1) \\
    & \times \int_{\eta_1}^\eta d\eta_2 \, \frac{j_2[k(\eta - \eta_2)]}{k^2 (\eta - \eta_2)^2} \frac{\partial \mathcal{T}(\eta_2,k)}{\partial \eta_2} \,,
\end{split}
\end{equation}
with $V(\eta_1,\eta_2)$ the visibility function, $j_n(x)$ a spherical Bessel function of the first kind, and $\mathcal{S}_\ell(\eta,k)= \frac{\ell + 2}{2\ell + 1} j_{\ell - 1}[k(\eta_0 - \eta)]  -  \frac{\ell-1}{2\ell + 1} j_{\ell + 1}[k(\eta_0 - \eta)]$. In our numerical implementation, we take only the portion of Eq.~(\ref{eq:Ph:constant}) which is super-horizon at collapse ($k < \eta_c^{-1}$) as a conservative estimate, as we do not track the sub-horizon evolution in detail. The visibility function and tensor transfer function are computed by \texttt{CLASS}~\cite{CLASS:I,CLASS:II}. Further details of the derivation can be found in Ch.~10 of Ref.~\cite{Rubakov:2011} or Sec.~3 of Ref.~\cite{Ireland:2025yqr}.

In Fig.~\ref{fig:B-mode_DW_r_0.001}, we show the resulting $B$-mode spectra for DW networks satisfying the various curvature perturbation bounds of Fig.~\ref{fig:vaOfetac_P_zeta}, together with the scale-invariant inflationary predictions for $r=10^{-3}$ and $r=0.036$. The CMB-constrained curves with the latest collapse times reach the largest amplitudes; for such DW networks, the $B$-mode amplitude can be comparable to or even exceed the inflationary signal for $r \sim 0.03$. The non-observation of such $B$-mode signals can therefore be used to constrain this scenario, along the lines of those constraints obtained in Ref.~\cite{Zebrowski:2026pye} for a broader class of post-inflationary sources. 

The spectral shape also depends on the DW collapse time. For the CMB-constrained benchmarks with later collapse, the spectra peak around the recombination bump at $\ell \sim 100$, just as the inflationary signal. Despite this similarity, the spectra are distinct, as the DW signal lacks the reionization bump at low $\ell$ and also falls off more quickly at high $\ell$. Earlier collapse tends to shift the peak towards higher multipoles, reaching $\ell \sim 1000$ for some of the benchmarks shown. These differences in amplitude and spectral shape may help to distinguish a DW-sourced signal from inflationary $B$ modes.

\section{Conclusion}

We have shown that finite-lifetime DW networks can source distinctive cosmological perturbations throughout their evolution. Using the USM, we derived analytic expressions for the scalar and tensor power spectra in the long-wavelength regime, finding a fixed ratio between their amplitudes. Unlike perturbations generated at the phase transition time when the wall forms, these spectra grow as the network evolves, driven by the increasing correlation length and the slower redshifting of the wall energy density relative to radiation. The spectra therefore peak at the collapse scale, with an amplitude that depends strongly on the DW lifetime.

Using various curvature perturbation constraints, we derived upper bounds on the allowed energy density in DWs at collapse. We demonstrated how such bounds could be translated into constraints on specific scenarios, focusing on two representative models: DWs from spontaneous $\mathbb{Z}_2$ breaking, and axion DWs. We further showed that, when curvature bounds are saturated, the corresponding tensor perturbations can produce CMB $B$-mode signals comparable to or larger than those from inflation with $r \sim 0.03$. Such signals may therefore be within reach of upcoming CMB polarization experiments.

\section{Acknowledgments}

We would like to thank Gordan Krnjaic for contributions at an earlier stage of this project. Claude and ChatGPT were used to cross check our calculation of the power spectrum, and Fig.~S-1 was generated using Claude. FY is supported by the Cluster of Excellence $\textit{PRISMA}^{++}$ (EXC 2118/2, Project ID 390831469), funded by the German Research Foundation (DFG). YT and FY are supported by the NSF Grant PHY-2412701. ANI is supported by NSF Grant PHY-2310429, Simons Investigator Award No.~824870, DOE HEP QuantISED award \#100495, the Gordon and Betty Moore Foundation Grant GBMF7946, and the U.S.~Department of Energy (DOE), Office of Science, National Quantum Information Science Research Centers, Superconducting Quantum Materials and Systems Center (SQMS) under contract No.~DEAC02-07CH11359. YT would also like to thank the host of Fermilab URA Visiting Scholar Program, supported by the URA-22-F-13 fund and the Tom and Carolyn Marquez Chair Fund for its generous support. YT would also like to thank the Aspen Center for Physics (supported by NSF grant PHY-2210452) and the Munich Institute for Astro-, Particle and BioPhysics
(funded by the DFG under Germany’s Excellence Strategy EXC-2094-390783311).

\bibliography{ref}

\clearpage
\onecolumngrid
\newpage

\widetext
 \begin{center}
   \textbf{\large SUPPLEMENTAL MATERIAL \\[.2cm] ``Cosmological Signals from Finite-Lifetime Domain Wall Networks''}\\[.2cm]
  \vspace{0.05in}
  {Fengwei Yang, Kylar Greene, Aurora Ireland, and Yuhsin Tsai}
\end{center}

\setcounter{equation}{0}
\setcounter{figure}{0}
\setcounter{table}{0}
\setcounter{page}{1}
\setcounter{section}{0} 
\makeatletter
\renewcommand{\thesection}{S-\Roman{section}}
\renewcommand{\theequation}{S-\arabic{equation}}
\renewcommand{\thefigure}{S-\arabic{figure}}

\section{Unequal-Time Correlators}

In this section, we demonstrate how to derive the unequal-time correlators for the Fourier modes of $T^{00}$ and $T^{ij}$. The induced metric on the wall worldsheet reads
\begin{equation}
    h_{ab} = a^2(\eta)
    \begin{pmatrix}
    -(1 - \dot{\bf x}^2) & 0 & 0 \\
    0 & {\bf x}'^{(1)^2} & 0 \\
    0 & 0  & {\bf x}'^{(2)^2} \\
    \end{pmatrix} \,, \quad h = -a^6(\eta)(1 - \dot{\bf x}^2) {\bf x}'^{(1)^2}{\bf x}'^{(2)^2}
    \,,  
\end{equation}
where $\dot{\bf x}=\frac{d{\bf x}}{d\eta}$, ${\bf{x}}'^{(b)} = \frac{d{\bf x}}{d\zeta^b},b=1,2$, and we consider the transverse gauge such that $\dot{\bf x}\cdot {\bf{x}}'^{(b)}=0$. We further assume ${\bf{x}}'^{(1)}\cdot {\bf{x}}'^{(2)}=0$ since in the Unconnected Segment Model (USM) the wall segment is featureless.

After substitution of the induced metric and integration over conformal time, the energy-momentum tensor in Eq.~(\ref{eq:Tmunu}) reads
\begin{eqnarray}
     T_{\mu\nu}(\eta, {\bf x}) = -a\sigma \int d^2\zeta \,\delta^{(3)}\left[ {\bf x} - {\bf x}_{\rm DW}(\eta,  \zeta^1,\zeta^2) \right] \Big\{ \varepsilon \dot{x}^{\mu} \dot{x}^{\nu} 
     - \varepsilon^{-1} \left( {\bf x}'^{(2)^{2}}{ x^{'(1)}}^{\mu} { x^{'(1)}}^{\nu} + {\bf x}'^{(1)^{2}}{ x^{'(2)}}^{\mu} { x^{'(2)}}^{\nu} \right) \Big\} \,,
\end{eqnarray}
where $\dot{x}^{\mu}=(1,\, \dot{\bf x} ),\,{x^{'(b)}}^{\mu}=(0,\,{\bf{x}}'^{(b)})$, and $\varepsilon$ is the coordinate energy per unit area
\bea
    \varepsilon \equiv \sqrt{\frac{{\bf x}^{'(1)^2} {\bf x}^{'(2)^2}}{1 - \dot{\bf x}^2}} \,,
\eea
which is related to the physical energy of the wall segment as 
\bea
    E=\sigma a^2\int d\zeta^1 d\zeta^2 \varepsilon \,.
\eea
By taking the spatial Fourier transform of the stress tensor $\Theta_{\mu \nu} (\eta, {\bf k}) = \int d^3 {\bf{x}} \,e^{-i \bf{k} \cdot \bf{x}}\, T_{\mu \nu} (\eta, {\bf x})$, we obtain
\bea
    \Theta_{\mu \nu}  (\eta, {\bf k}) = -a\sigma \!\int\! d^2 \zeta \, e^{-i {\bf k} \cdot {\bf x}_{\rm DW} } \bigg[ \varepsilon\, \dot{x}^{\mu} \dot{x}^{\nu} 
     - \varepsilon^{-1}  \big( {\bf x}'^{(2)^{2}}{ x^{'(1)}}^{\mu} { x^{'(1)}}^{\nu} + {\bf x}'^{(1)^{2}}{ x^{'(2)}}^{\mu} { x^{'(2)}}^{\nu}  \big) \bigg] \,.
\eea
We parameterize the DW trajectory as ${\bf{x}}_{\rm DW}(\zeta^a) = {\bf{x}}_0 + v \eta \hat{\dot{\bf{x}}} + \zeta^1 \hat{\bf{x}}^{'(1)} + \zeta^2 \hat{\bf{x}}^{'(2)}$.

\begin{figure}[t!]
    \centering
    \includegraphics[width=0.85\linewidth]{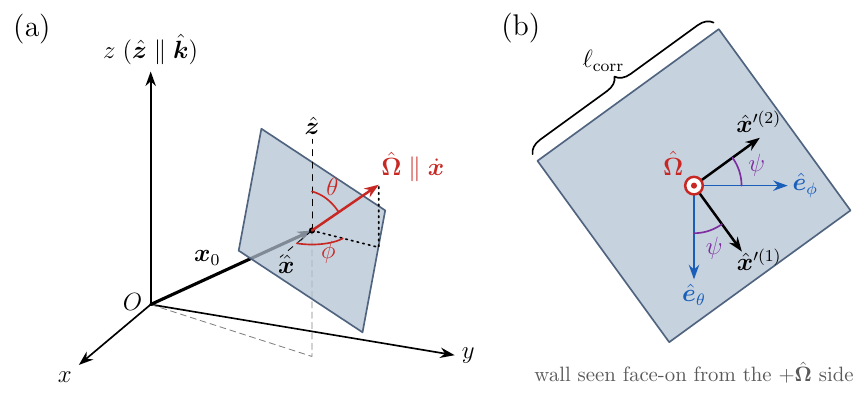}
    \caption{The worldsheet coordinates of a flat square DW segment modeled in the USM.} 
    \label{fig:coordinates}
\end{figure}

After performing the integral over $\zeta$, the $00$- and $ij$-components of the energy-momentum tensor are 
\begin{eqnarray}
    \Theta_{00} (\eta, {\bf k}) &=&-a\sigma\gamma e^{-i {\bf k} \cdot ({\bf x}_0 + v \eta \dot{\hat{\bf x}})}\frac{2 \sin(\frac{k\ell_{\rm corr}}{2} \hat{x}_z^{'(1)})}{k\hat{x}_z^{'(1)}}  \frac{2 \sin(\frac{k\ell_{\rm corr}}{2} \hat{x}_z^{'(2)})}{k\hat{x}_z^{'(2)}} \,,~~~\label{eq:Theta00}\\
    \Theta_{ij} (\eta, {\bf k}) &=& \Theta_{00} (\eta, {\bf k}) [v^2 \hat{\dot x}_i \hat{\dot x}_j - \gamma^{-2} (\hat{x}^{'(1)}_i\hat{x}^{'(1)}_j + \hat{x}^{'(2)}_i\hat{x}^{'(2)}_j )] \,,~~\label{eq:Thetaij}
\end{eqnarray}
where $\gamma=(1-v^2)^{-1/2}$ is the Lorentz factor. The coordinates of the DW segment are shown in Fig.~\ref{fig:coordinates}, where we set $\hat{\bf k} \parallel \hat{\bf z}$ 
without loss of generality. 
$\hat{x}_i^{'(b)}$ and $\hat{\dot x}_i$ are the $i$-th component of the unit vector of the wall tangent vector ${\bf x}^{'(b)}$ and the wall velocity $ \dot{\bf x}$ respectively, with $i=x,\,y,\,z$ labeling Cartesian coordinates. The unit vectors can be expressed in terms of three angles $\{\theta,\phi,\psi\}$ that fix the DW segment orientation as shown in Fig.~\ref{fig:coordinates}:
\begin{equation}
    \hat{\boldsymbol{\Omega}}=\hat{\dot{\bf{x}}} = 
\begin{pmatrix}
\sin\theta\cos\phi \\
\sin\theta\sin\phi  \\
\cos\theta \\
\end{pmatrix}, ~\hat{\bf{x}}^{'(1)}=\begin{pmatrix}
\cos\phi\cos\theta\cos\psi-\sin\phi\sin\psi \\
\sin\phi\cos\theta\cos\psi+\cos\phi\sin\psi  \\
-\sin\theta\cos\psi \\
\end{pmatrix},
\end{equation}
and
\begin{equation}
\hat{\bf{x}}^{'(2)}=\begin{pmatrix}
-\cos\phi\cos\theta\sin\psi-\sin\phi\cos\psi \\
-\sin\phi\cos\theta\sin\psi+\cos\phi\cos\psi  \\
\sin\theta\sin\psi \\
\end{pmatrix}.    
\end{equation}

The power spectrum of cosmological perturbations is sourced by the unequal-time correlator (UETC)
\begin{equation}\label{eq:uetc_def}
    \langle \Theta_{\mu \nu} (\eta_1,{\bf k}) \Theta_{\mu \nu}^* (\eta_2, {\bf k}') \rangle' \equiv \Theta_{\mu\nu,\mu\nu}(\eta_1, \eta_2, k)\,,
\end{equation}
where the $'$ on the correlator denotes that momentum-conserving factors $(2\pi)^3\delta^{(3)} ({\bf k}-{\bf k}')$ have been stripped off. Note that in $\langle \cdots \rangle$ there is no sum over the indices $\mu,\nu$, and the ensemble average is taken over all possible DW configurations, including their orientation, velocity direction, and initial positions within a comoving volume $V = \int d^3{\bf x}_0$, where the wall segments are distributed. 

In the USM, the ensemble average in Eq.~(\ref{eq:uetc_def}) can be performed analytically
\begin{equation}
\label{eq:ensemble_ave}
\displaystyle{\int}\frac{d^3{\bf x}_0}{V} \int\frac{d^2\hat{\boldsymbol{\Omega}}}{4\pi} \,\int_0^{2\pi} \frac{d\psi}{2\pi} \, \Theta_{\mu \nu}(\eta_1,{\bf k}) \Theta_{\mu \nu}^*(\eta_2,{\bf k}')\,,
\end{equation}
where the stress-energy tensor of individual DW segment in \cref{eq:Theta00,eq:Thetaij} is used and the scaling property of the DW network is imposed such that the comoving number density of the DW segment is one per correlation volume, i.e., $1/V=1/\max(\xi\eta_1,\xi\eta_2)^{3}$.  
See Fig.~\ref{fig:coordinates} for the  definition of ${\bf x}_0$, $\hat{\boldsymbol{\Omega}}$, and $\psi$. 

\begin{figure}[t!]
        \centering
\includegraphics[width=0.5\linewidth]{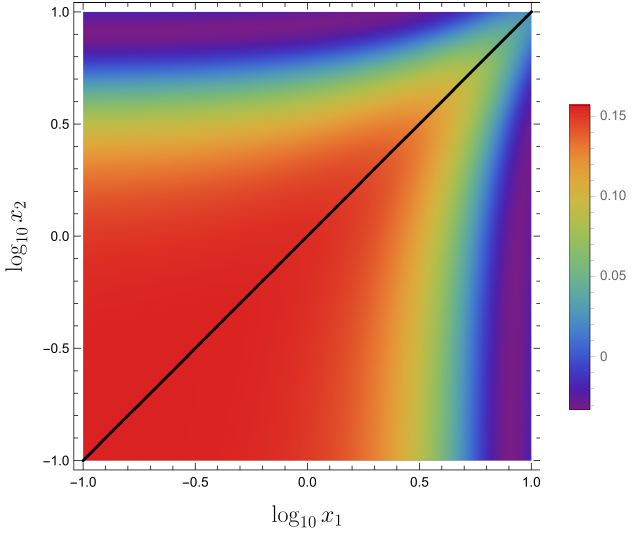}
    \caption{The UETC of $00$-component of the Fourier transform of the energy-momentum tensor as a function of $x_1$ and $x_2$, where $x_{1,2}\equiv k\eta_{1,2}$ is a dimensionless time variable. The color represents the amplitude of the rescaled dimensionless UETC $\Theta_{00,00}(\frac{x_1}{k},\frac{x_2}{k},k)/(\sigma^2\gamma^2a_1 a_2\sqrt{\eta_1\eta_2})$, with $\Theta_{00,00}$ in Eq.~(\ref{appeq:Theta0000}) and $a_{1,2}=a(\eta_{1,2})$.} 
    \label{fig:uetc0000}
\end{figure}

Combining \cref{eq:Theta00,eq:Thetaij,eq:uetc_def,eq:ensemble_ave}, one can derive the UETC of the 00- and $ij$-components
\begin{eqnarray}
\label{appeq:Theta0000}
\Theta_{00,00} &=& \frac{a_1a_2}{\max(\ell_{\rm corr_1},\ell_{\rm coll_2})^{3}} \frac{4\sigma^2\gamma^2}{\pi k^4} \int \frac{d\cos\theta}{\sin^{4}\theta} \frac{d\psi}{\sin^{2}(2\psi)} e^{iv k\eta_-\cos\theta }\left[\cos\left(\frac{k\xi\eta_-}{2}\sin\theta\cos\psi\right) - \cos\left(\frac{k\xi\eta_+}{2}\sin\theta\cos\psi\right)\right]\nonumber\\
    &\times& \left[\cos\left(\frac{k\xi\eta_-}{2}\sin\theta\sin\psi\right) - \cos\left(\frac{k\xi\eta_+}{2}\sin\theta\sin\psi\right)\right] \,,~~
\end{eqnarray}
\begin{eqnarray}
\label{appeq:Theta1212}
    \Theta_{ij,ij}&=& \frac{a_1a_2}{\max(\ell_{\rm corr_1},\ell_{\rm coll_2})^{3}} \frac{2\sigma^2\gamma^2}{\pi^2k^4} \int \frac{d\cos\theta}{\sin^{4}\theta} \frac{d\psi}{\sin^{2}(2\psi)}d\phi e^{iv k\eta_-\cos\theta }\left[\cos\left(\frac{k\xi\eta_-}{2}\sin\theta\cos\psi\right) - \cos\left(\frac{k\xi\eta_+}{2}\sin\theta\cos\psi\right)\right]\nonumber\\
    &\times& \left[\cos\left(\frac{k\xi\eta_-}{2}\sin\theta\sin\psi\right) - \cos\left(\frac{k\xi\eta_+}{2}\sin\theta\sin\psi\right)\right]
   F_{ij,ij}(v,\theta,\phi,\psi) \,,
\end{eqnarray}
where the UETCs are functions of $\eta_1,\eta_2,k$, and we introduce the shorthand notations $a_i=a(\eta_i)$, $\ell_{\rm coll_i}=\ell_{\rm coll}(\eta_i)$, $\eta_\pm \equiv \eta_1 \pm \eta_2$, and $F_{ij,kl}(v,\theta,\phi,\psi)\equiv[v^2 \hat{\dot x}_i \hat{\dot x}_j - (1-v^2) (\hat{x}^{'(1)}_i\hat{x}^{'(1)}_j + \hat{x}^{'(2)}_i\hat{x}^{'(2)}_j )][v^2 \hat{\dot x}_k \hat{\dot x}_l - (1-v^2) (\hat{x}^{'(1)}_k\hat{x}^{'(1)}_l + \hat{x}^{'(2)}_k\hat{x}^{'(2)}_l )]$. We show $\Theta_{00,00}(\eta_1,\eta_2,k)$ as a function of $ k\eta_1$ and $k\eta_2$ in Fig.~\ref{fig:uetc0000}. The correlation peaks along the diagonal, with off-diagonal contributions suppressed. We therefore approximate the DW-induced curvature power spectrum~\eqref{eq:Pzeta:EMtensor:relation}, which depends only on the $\Theta_{00,00}$ component, using the equal-time correlator (ETC). For the tensor power spectrum, on the other hand, we retain the full time integration with the UETC.

\section{Derivation of Curvature Perturbation Power Spectrum}

For a superhorizon mode $k$, the curvature perturbation evaluated at horizon reentry is given by
\begin{equation}
    \zeta=-\Psi-H\frac{\delta\rho_{\rm DW}}{d{\bar{\rho}}_{\rm tot}/dt} \,,
\end{equation}
where $\Psi$ is the gravitational potential, $\delta\rho_{\rm DW}(\eta,{\bf x})=\rho_{\rm DW}(\eta,{\bf x})-\bar{\rho}_{\rm DW}$ is the energy density fluctuation of domain walls, and $\bar\rho_{\rm tot}$ is the total energy density, which we take to be $\bar\rho_{\rm tot} \gg \delta\rho_{\rm DW}$. We choose the spatially flat gauge such that $\Psi=0$. From the continuity equation 
\begin{equation}
    \frac{d{\bar{\rho}}_{\rm tot}}{dt}=-3H(1+w)\bar{\rho}_{\rm tot} \,,
\end{equation}
we thus have during radiation domination
\begin{equation}
\label{eq:zeta_RD}
    \zeta(\eta,{\bf x})= \frac{\delta\rho_{\rm DW}(\eta,{\bf x})}{4\bar{\rho}_{\rm tot}(\eta)} \,.
\end{equation}
The Fourier transform of this quantity for $k>0$ is
\bea\label{eq:zeta_RD_FT}
    \zeta_{\mathbf{k}}(\eta) = \frac{a^{-2}}{4 \bar{\rho}_{\rm tot}} \Theta_{00}(\eta,\mathbf{k}) \,,
\eea
where the scale factors come from the convention for the energy density $\rho_{\rm DW} = a^{-2} T_{00}$. The dimensionless curvature power spectrum $\mathcal{P}_\zeta(\eta,k)$ is defined as
\bea
    \langle \zeta_{\mathbf{k}}(\eta) \zeta_{\mathbf{k}'}^*(\eta) \rangle = (2\pi)^3 \delta^{(3)}(\mathbf{k} - \mathbf{k}') \frac{2\pi^2}{k^3} \mathcal{P}_\zeta(\eta,k) \,.
\eea
Using Eq.~(\ref{eq:zeta_RD_FT}) and the definition of the UETC in Eq.~(\ref{eq:uetc_def}), this quantity thus reads
\begin{equation}\label{eq:Pzeta:EMtensor:relation}
    \mathcal{P}^{\rm DW}_\zeta(\eta,k) =\frac{k^3}{2\pi^2}\frac{a^{-4}}{16\bar{\rho}^2_{\rm tot}}\Theta_{00,00}(\eta,\eta,k) \,.
\end{equation}
Evaluating Eq.~(\ref{appeq:Theta0000}) at equal times $\eta_1 = \eta_2 = \eta$ and taking the long-wavelength limit $(\xi k \eta) \ll 1$, the angular integrals in $\Theta_{00,00}(\eta,\eta,k)$ can be performed trivially and one obtains
\bea
    \Theta_{00,00}(\eta,\eta,k) \big|_{\xi k \eta \ll 1} \simeq a^2 \sigma^2 \gamma^2 \xi \eta \,.
\eea
Substituting into Eq.~(\ref{eq:Pzeta:EMtensor:relation}) and using that the average DW energy density is $\bar{\rho}_{\rm DW} = \sigma/(\xi a \eta)$, the power spectrum in the long-wavelength regime becomes
\bea\label{eq:Pzeta:DW:arbitrary:eta}
    \mathcal{P}^{\rm DW}_\zeta(\eta,k) =\frac{\gamma^2}{32\pi^2} f_{\rm DW}^2 (\xi k \eta)^3  \,,
\eea
where we have used the definition $f_{\rm DW} = \bar{\rho}_{\rm DW}/\bar{\rho}_{\rm tot}$. During RD, the fractional energy density in DWs is related to its initial value as
\bea
    f_{\rm DW} = f_{\rm DW}^i \left( \frac{g_\star}{g_{\star,i}} \right)^{1/3} \left( \frac{a}{a_i} \right)^3 \left( \frac{\eta}{\eta_i} \right)^{-1} \,,
\eea
where $g_{\star,i} \equiv g_\star(T_i)$ and we have approximated $g_{\star,s} \simeq g_\star$. Using this relation in Eq.~(\ref{eq:Pzeta:DW:arbitrary:eta}) and converting the scale factor to conformal time using the RD era relation $(a/a_i) = (g_\star/g_{\star,i})^{-1/6} (\eta/\eta_i)$, we finally obtain the expression for the scalar power spectrum cited in the main text
\bea\label{appeq:Pzeta}
    \mathcal{P}_\zeta^{\rm DW}(\eta,k) =\frac{\gamma^2}{32\pi^2} (f^i_{\rm DW})^2 (\xi k\eta_i)^3 \left( \frac{g_\star}{g_{\star,i}} \right)^{-1/3} \left(\frac{\eta}{\eta_i} \right)^7 \,.
\eea


\section{Derivation of Tensor Perturbation Power Spectrum}

\subsection{General Formalism}

We begin with the perturbed FLRW metric in conformal coordinates, 
\begin{equation}\label{eq:flrw} 
    ds^2 = a^2(\eta) \left[ -d\eta^2 + (\delta_{ij} + h_{ij}) \, dx^i dx^j \right] \,, 
\end{equation} 
where $h_{ij}$ denotes the spatial metric perturbation. We work in a gauge in which the temporal components vanish, $h_{0\mu} = 0$, and further impose the transverse-traceless (TT) condition on the spatial components, $\partial^i h_{ij} = 0$, $h_i^i = 0$. At linear order, the tensor perturbations satisfy the wave equation
\begin{equation}\label{eq:eom_h}
    \ddot{h}_{ij} + 2 \mathcal{H} \dot{h}_{ij} - \nabla^2 h_{ij} \; = \; 16\pi G a^2 \Pi_{ij}^{\text{TT}} \,,
\end{equation}
where $\dot{\textcolor{white}{x}}$ denotes a derivative with respect to conformal time, $\mathcal{H}=\dot{a}/a$ is the conformal Hubble parameter, and $\Pi_{ij}^{\rm TT}$ is the transverse-traceless part of the anisotropic stress. To solve, we first decompose the metric perturbation into Fourier modes
\begin{equation}\label{eq:fourier}
    h_{ij}(\eta,\mathbf{x}) = \int \!\frac{d^3k}{(2\pi)^3} \, h_{ij}(\eta,\mathbf{k}) \, e^{i\mathbf{k} \cdot \mathbf{x}} \,.
\end{equation}
In Fourier space, Eq.~\eqref{eq:eom_h} becomes
\begin{equation}\label{eq:eom_h_fourier}
    \ddot{h}_{ij}(\eta,\mathbf{k}) + 2 \mathcal{H} \dot{h}_{ij}(\eta,\mathbf{k}) + k^2 h_{ij}(\eta,\mathbf{k}) = 16\pi G a^2 \Pi_{ij}^{\text{TT}}(\eta,\mathbf{k}) \,. 
\end{equation}
Introducing the dimensionless variable $y\equiv k\eta$, this equation can equivalently be written
\bea\label{eq:eom_hy}
    {h}''_{ij}(\eta,\mathbf{k}) + 2 \frac{\mathcal{H} }{k}{h}'_{ij}(\eta,\mathbf{k}) +h_{ij}(\eta,\mathbf{k}) = \frac{16\pi G a^2 }{k^2} \Pi_{ij}^{\text{TT}}(\eta,\mathbf{k}) \,,
\eea
where $'$ denotes a derivative with respect to $y$. The general solution can be found by the method of Green's functions. We write
\begin{equation}\label{eq:hij:gen:sol}
    {h}_{ij}(y,\mathbf{k})=\frac{16\pi G}{k^2} \int_{y_i}^y \!dz \,\mathcal{G}(y,z)\,a^2(z)\, \Pi_{ij}^{\text{TT}}(z,\mathbf{k}) \,,
\end{equation}
where $y_i = k\eta_i$, $z = k \eta'$ is a dimensionless integration variable, and the functions $a(z)$ and $\Pi_{ij}^{\text{TT}}(z,\mathbf{k})$ are understood to be evaluated at conformal time $\eta' = z/k$. The Green's function satisfies
\begin{equation}\label{eq:Greens:function}
    \mathcal{G}''(y,z)+ 2 \frac{\mathcal{H}}{k}\mathcal{G}'(y,z) +\mathcal{G}(y,z)= \delta(y-z) \,.
\end{equation}

To proceed further, note that the TT conditions imply that each spatial tensor perturbation has two degrees of freedom for each wavevector $\mathbf{k}$, allowing us to decompose into polarization modes
\begin{equation}
    h_{ij}(\eta,\mathbf{k}) = \sum_{\lambda = +, \times} h_{\lambda}(\eta,\mathbf{k})\, \varepsilon_{ij}^{\lambda}(\mathbf{k}) \,.
\end{equation}
The polarization tensors $\varepsilon_{ij}^{\lambda}(\mathbf{k})$ form an orthonormal basis for symmetric, transverse-traceless tensors, and satisfy $k^i \varepsilon^\lambda_{ij} = 0$, $\delta^{ij} \varepsilon^\lambda_{ij} = 0$, and $\varepsilon_{ij}^\lambda \varepsilon_{ij}^{\lambda'} = \delta_{\lambda \lambda'}$. To construct the basis, we introduce two orthonormal unit vectors ${\bf u}(\mathbf{k})$ and ${\bf v}(\mathbf{k})$, both perpendicular to $\mathbf{k}$. The polarization tensors are then 
\begin{equation}
    \varepsilon_{ij}^+(\mathbf{k}) = \frac{1}{\sqrt{2}} \left[ u_i(\mathbf{k}) u_j(\mathbf{k}) - v_i(\mathbf{k}) v_j(\mathbf{k}) \right] \,, \quad \varepsilon_{ij}^\times(\mathbf{k}) = \frac{1}{\sqrt{2}} \left[ u_i(\mathbf{k}) v_j(\mathbf{k}) + v_i(\mathbf{k}) u_j(\mathbf{k}) \right] \,.
\end{equation}
The polarization amplitudes are recovered through
\bea\label{eq:hlambda:hij:relation}
    h_\lambda(\eta,\mathbf{k}) = \varepsilon_{ij}^\lambda(\mathbf{k}) h_{ij}(\eta,\mathbf{k}) \,.
\eea
Meanwhile, the transverse-traceless part of the anisotropic stress $\Pi_{ij}^{\rm TT}$ can be decomposed as
\begin{equation}\label{eq:PiTT}
    \Pi_{ij}^{\mathrm{TT}}(\eta,\mathbf{k}) = a^{-2}(\eta)\,\Lambda_{ij,kl}(\mathbf{k})\,
     \Theta_{kl}(\eta,\mathbf{k}) \,, 
\end{equation}
where $\Theta_{\mu \nu}$ is the Fourier transform of the stress tensor introduced before, $\Theta_{\mu \nu} (\eta, {\bf k}) = \int d^3 {\bf{x}} \,e^{-i \bf{k} \cdot \bf{x}}\, T_{\mu \nu} (\eta, {\bf x})$, and
\begin{equation}
   \Lambda_{ij,kl}(\mathbf{k}) = P_{ik}P_{jl}-\tfrac{1}{2}P_{ij}P_{kl} \,, \quad \text{with} \quad
   P_{ij}(\mathbf{k}) = \delta_{ij}-\hat{k}_i\hat{k}_j \,.
\end{equation}
Note that this decomposition makes it clear that $k^i\Pi_{ij}^{\rm TT} = 0$ and $\delta^{ij} \Pi_{ij}^{\rm TT} = 0$. Combining Eqs.~(\ref{eq:hij:gen:sol}), (\ref{eq:hlambda:hij:relation}), and (\ref{eq:PiTT}), one can write the two-point function of tensor amplitudes as
\bea\label{eq:hlambda:2pt}
    \langle h_\lambda(\eta,\mathbf{k}) h_{\lambda'}^*(\eta,\mathbf{k}') \rangle = \left(\frac{16\pi G}{k^2}\right)^2\int_{y_i}^{y} dy' \int_{y_i}^{y} dy'' \, \mathcal{G}(y, y')  \, \mathcal{G}(y, y'') \varepsilon_{\lambda}^{ij} \varepsilon_{\lambda'}^{*kl} \langle \Theta_{ij} (y'/k,\mathbf{k}) \Theta_{kl}^* (y''/k',\mathbf{k}')\rangle \,.
\eea
Using the definition of the UETC in Eq.~(\ref{eq:uetc_def}) as well as the definition of the power spectrum, we find 
\bea\label{eq:psh}
    P_h^\lambda = \left(\frac{16\pi G}{k^2}\right)^2\int_{y_i}^{y} dy' \int_{y_i}^{y} dy'' \, \mathcal{G}(y, y')  \, \mathcal{G}(y, y'')  \varepsilon_{\lambda}^{ij} \varepsilon_{\lambda}^{*kl} \Theta_{ij,kl}\left( \frac{y'}{k}, \frac{y''}{k},k\right) \,.
\eea
The total tensor power spectrum is given by the polarization sum 
\bea\label{eq:Ph}
    P_h(\eta,k) = \sum_{\lambda = +,\times} P_h^\lambda \,,
\eea
and the dimensionless version is
\begin{equation}\label{eq:dimless:version}
    \mathcal{P}_h(\eta,k) = \frac{k^3}{2\pi^2} P_h(\eta,k) \,.
\end{equation}

\subsection{Tensor Perturbations Sourced by DWs in the Scaling Regime}

Let us now specialize to our case of interest. In evaluating Eq.~(\ref{eq:psh}) for the DW network, the lower limit of integration $y_i = k\eta_i$ corresponds to DW formation, when the network can start sourcing tensor modes, while
the upper limit $y=k\eta$ is taken to be some time before collapse, $\eta \leq \eta_c$. During RD, the solution to Eq.~(\ref{eq:Greens:function}) is given by
\begin{equation}
    \mathcal{G}(y, z)= \frac{z\sin(y-z)}{y} \mathsf{H}(y-z) \,,
\end{equation}
where $\mathsf{H}(y)$ is the Heaviside function. Performing the polarization sum in Eq.~\eqref{eq:Ph}, we find that 
\begin{equation}\label{eq:pol:sum}
    \sum_{\lambda = +, \times} \varepsilon_{\lambda}^{ij} \varepsilon_{\lambda}^{*kl} \Theta_{ij,kl}=2\Theta_{12,12}+\frac{1}{2}(\Theta_{11,11}+\Theta_{22,22}-2\Theta_{11,22}) \,.
\end{equation}
Evaluating Eq.~(\ref{appeq:Theta1212}) at equal times $\eta_1 = \eta_2 = \eta$ and taking the long-wavelength limit $(\xi k \eta) \ll 1$, we obtain
\begin{eqnarray}
\label{appeq:Thetaijij}
    &\Theta_{11,11}(\eta,\eta,k) \big|_{\xi k \eta \ll 1} = \Theta_{22,22}(\eta,\eta,k)\big|_{\xi k \eta \ll 1} \simeq \frac{1}{15} a^2 \sigma^2 \gamma^2 \xi \eta (8-20v^2+15v^4) \,,\\
    &\Theta_{11,22}(\eta,\eta,k)\big|_{\xi k \eta \ll 1} \simeq \frac{1}{15} a^2 \sigma^2 \gamma^2 \xi \eta (6-20v^2+15v^4) \,,\\
    &\Theta_{12,12}(\eta,\eta,k)\big|_{\xi k \eta \ll 1} \simeq \frac{1}{15} a^2 \sigma^2 \gamma^2 \xi \eta \,.
\end{eqnarray}
Inserting the above into Eq.~(\ref{eq:pol:sum}), we find
\begin{equation}\label{eq:sum:1212}
    \sum_{\lambda = +, \times} \varepsilon_{\lambda}^{ij} \varepsilon_{\lambda}^{*kl} \Theta_{ij,kl} = 4 \Theta_{12,12} \,.
\end{equation}
Note that, other than the velocity dependence in the Lorentz factor, all explicit velocity dependence has canceled between the components. Also note that although Eq.~(\ref{eq:sum:1212}) was derived in the equal-time limit, the same relation holds for the UETC at leading order in the long-wavelength expansion, as can be verified by expanding the cosines in Eq.~(\ref{appeq:Theta1212}). The DW tensor power spectrum is therefore 
\begin{equation}
    P_h^{\rm DW} \!\left(y,k\right) = 2 \left( \frac{16\pi G}{k^2} \right)^2  I_y(y,k) \,,
\end{equation}
where we have defined
\bea\label{eq:Iy:def}
    I_y(y,k) = \frac{1}{y^2} \int_{y_i}^{y} \! dy_1 \int_{y_i}^y \! dy_2 \, y_1 y_2 \big( \cos(y_1-y_2) - \cos(2y - y_1-y_2) \big) \Theta_{12,12} \left( \frac{y_1}{k}, \frac{y_2}{k}, k \right) \,.
\eea
In the long-wavelength limit, the angular integrals in the expression for $\Theta_{12,12}$ in Eq.~(\ref{appeq:Theta1212}) can be straightforwardly evaluated to yield
\bea\label{eq:Theta1212:expansion}
    \Theta_{12,12} \left( \frac{y_1}{k}, \frac{y_2}{k}, k \right) \big|_{y_1,y_2 \ll 1} \simeq \frac{1}{15} \frac{ (y_1 y_2)^2}{\max(y_1,y_2)^3} a_1 a_2 \sigma^2 \gamma^2 \xi k^{-1} \,.
\eea
Also expanding the difference of cosines in Eq.~(\ref{eq:Iy:def}) for $y_{1,2} \ll 1$, approximating the scale factor during RD by $a_{1,2}/a \simeq y_{1,2}/y$, and taking the lower limits of integration to zero $y_{1,2} \rightarrow 0$, the double time integral evaluates to
\bea
    I_y \simeq \frac{19}{56700} a^2 \sigma^2\gamma^2 \xi \, y^5 k^{-1} \,.
\eea
We thus obtain for the super-horizon tensor power spectrum
\begin{equation}\label{appeq:Ph}
    P_h^{\rm DW}(\eta,k) \simeq \frac{2432}{14175} \pi^2 G^2 a^2 \sigma^2 \gamma^2 \xi \eta^5 \,,
\end{equation}
which indeed displays the expected white noise scaling $P_h \propto k^0$. Multiplying by $k^3/2\pi^2$ to obtain the dimensionless version, using the RD era relation $(a/a_i) = (g_\star/g_{\star,i})^{-1/6} (\eta/\eta_i)$, and expressing the result in terms of the initial DW energy fraction $f_{\rm DW}^i$, we finally obtain
\begin{equation}\label{eq:Ph:386}
   \mathcal{P}_h^{\rm DW}(\eta,k) = 0.386 \cdot \frac{\gamma^2}{32\pi^2} (f^i_{\rm DW})^2 (\xi k\eta_i)^3 \left( \frac{g_\star}{g_{\star,i}} \right)^{-1/3} \left(\frac{\eta}{\eta_i} \right)^7 \,.
\end{equation}
Comparing with the scalar power spectrum in Eq.~(\ref{appeq:Pzeta}), we see that they are indeed directly proportional in the super-horizon regime, $\mathcal{P}_h^{\rm DW}(\eta,k) = 0.386 \mathcal{P}_\zeta^{\rm DW}(\eta,k)$.

\subsection{Enhancement of Super-Horizon Modes after DW Collapse}

Let us now consider the evolution of super-horizon tensor modes after the DW network collapses at $\eta_c$. For times $\eta > \eta_c$, the source vanishes and so the tensor mode equation of motion becomes
\bea
    \ddot{h}_\lambda + \frac{2}{\eta} \dot{h}_\lambda + k^2 h_\lambda = 0 \,.
\eea
The general solution reads
\bea\label{eq:h:postcollapse:gensol}
    h_\lambda(\eta,k) = A_\lambda(k) j_0(k\eta) + B_\lambda(k) y_0(k\eta) \,,
\eea
where $j_0(z) = \sin (z)/z$ and $y_0(z) = - \cos(z)/z$ are spherical Bessel functions of the first and second kind. The coefficients $A_\lambda$ and $B_\lambda$ can be fixed by matching this solution (and its first derivative) to the corresponding sourced solution (and its first derivative) evaluated at $\eta_c$. The solution after collapse, obtained by integrating the source up to $\eta_c$, is given by 
\bea
    h_\lambda(\eta,k) = \frac{16 \pi G}{k^2} \frac{1}{k\eta} \int_{y_i}^{y_c}\! dz \, z \sin(y-z) \varepsilon_\lambda^{ij} \Theta_{ij}\left( z/k, \mathbf{k} \right) \,,
\eea
where as before $y = k \eta$ and we have used the RD era Green's function in Eq.~(\ref{eq:Greens:function}). Using the identity $\sin(y-z) = \sin(y) \cos(z) - \cos(y) \sin(z)$ and demanding equality with Eq.~(\ref{eq:h:postcollapse:gensol}), it is straightforward to identify the coefficients
\bea
    A_\lambda(k) = \frac{16 \pi G}{k^2} \int_{y_i}^{y_c} \! dz \, z \cos(z) \varepsilon_\lambda^{ij} \Theta_{ij}\left( z/k, \mathbf{k} \right) \,,
\eea
and
\bea
    B_\lambda(k) = \frac{16 \pi G}{k^2} \int_{y_i}^{y_c} \! dz \, z \sin(z) \varepsilon_\lambda^{ij} \Theta_{ij}\left( z/k, \mathbf{k} \right) \,.
\eea
In the super-horizon regime $k\eta \ll 1$, we have that $j_0(k\eta) \simeq 1$ while $y_0(k\eta) \simeq - 1/k\eta$, and so we identify the $A_\lambda$ term as the constant mode and the $B_\lambda$ term as the decaying mode. More precisely, in the limit $k\eta_c \ll 1$ one obtains for the constant mode
\bea
    A_\lambda(k) \simeq h_{\lambda,c} + \eta_c \dot{h}_{\lambda,c} \,,
\eea
where $h_{\lambda,c} = h_\lambda(\eta_c)$ and $\dot{h}_{\lambda,c} = \dot{h}_\lambda |_{\eta_c}$ are the tensor amplitude and its derivative evaluated at collapse. As we will now see, this additional contribution $\eta_c \dot{h}_{\lambda,c}$ will give rise to a large enhancement of the late-time power spectrum relative to that at collapse. 

Let us neglect the correction from the decaying mode and proceed to evaluate the power spectrum of the constant super-horizon mode, which reads
\bea
    P_{h,\text{const}}^{\rm DW}(k) = \left( \frac{16 \pi G}{k^2} \right)^2 \int_{y_i}^{y_c} \! dz_1 \int_{y_i}^{y_c} \! dz_2 \, z_1 z_2 \cos(z_1) \cos(z_2) \sum_{\lambda = +,\times} \varepsilon_\lambda^{ij} \varepsilon_\lambda^{* kl} \Theta_{ij,kl}\bigg( \frac{z_1}{k}, \frac{z_2}{k}, k \bigg) \,. 
\eea
As before we use that at leading order in the long-wavelength expansion $\sum_{\lambda = +,\times} \varepsilon_\lambda^{ij} \varepsilon_\lambda^{* kl} \Theta_{ij,kl}(z_1/k, z_2/k, k) \simeq 4 \Theta_{12,12}(z_1/k, z_2/k, k)$, with $\Theta_{12,12}$ given by Eq.~(\ref{eq:Theta1212:expansion}). We thus write
\bea\label{eq:PA:inprogress}
    P_{h,\text{const}}^{\rm DW}(k) = 2 \left( \frac{16 \pi G}{k^2} \right)^2 I_{\rm const}(y_c) \,,
\eea
with
\bea
    I_{\rm const}(y_c) = \frac{2}{15} \sigma^2 \gamma^2 \xi k^{-1} \int_{y_i}^{y_c} \! dz_1 \int_{y_i}^{y_c} \! dz_2 \, \frac{z_1^3 z_2^3 a_1 a_2}{\text{max}(z_1,z_2)^3} \cos(z_1) \cos(z_2) \,.
\eea
As before, we approximate the scale factor by $a_{1,2}/a_c = z_{1,2}/y_c$ and take the lower limits of integration to zero, which yields
\bea
    I_{\rm const}(y_c) \simeq \frac{2}{15} \frac{a_c^2 \sigma^2 \gamma^2 \xi}{k y_c^2} \int_{0}^{y_c} \! dz_1 \int_{0}^{y_c} \! dz_2 \, \frac{z_1^4 z_2^4}{\text{max}(z_1,z_2)^3} \cos(z_1) \cos(z_2) \,.
\eea
Due to the symmetry of the integrand and integration limits, this integral can be evaluated analytically, resulting in
\bea
    I_{\rm const}(y_c) = \frac{2}{15} \frac{a_c^2 \sigma^2 \gamma^2 \xi}{k} y_c^5 f(y_c) \,,
\eea
where
\bea\label{eq:f:def}
    f(y_c) = \frac{4}{5} \left( \frac{1}{y_c^2} - \frac{10}{y_c^4} \right) - \frac{1}{4} \left( \frac{2}{y_c^2} - \frac{50}{y_c^4} + \frac{171}{y_c^6} \right) \cos(2 y_c) + \frac{1}{8} \left( \frac{26}{y_c^3} - \frac{246}{y_c^5} + \frac{171}{y_c^7} \right) \sin(2y_c) \,.
\eea
Substituting into Eq.~(\ref{eq:PA:inprogress}), we thus obtain for the super-horizon constant mode power spectrum
\bea
    P_{h,\text{const}}^{\rm DW}(k) = \frac{1024}{15} \pi^2 G^2 a_c^2 \sigma^2 \gamma^2 \xi \eta_c^5 f(k \eta_c) \,.
\eea
In order to quantify how much this power spectrum is enhanced relative to the value at collapse, we define the enhancement factor
\bea\label{eq:enhancement}
    \mathcal{E}(k \eta_c) \equiv \frac{P_{h,\text{const}}^{\rm DW}(k)}{P_h^{\rm DW}(\eta_c, k)} = \frac{7560}{19} f(k \eta_c) \,,
\eea
with $P_h^{\rm DW}$ given in Eq.~(\ref{appeq:Ph}). Note that the function $f$ in Eq.~(\ref{eq:f:def}) varies between $f(y_c \rightarrow 0) = 2/35$ for long-wavelength modes in the strict $k \rightarrow 0$ limit to $f(y_c = 1) \simeq 0.027$ for modes near the peak, though this latter limit should be understood as an estimate since this expression was derived using the long-wavelength approximation for the UETC, which is formally valid only for $k \eta_c \ll 1$. The enhancement factor then varies between $\mathcal{E} \simeq 22.7$ for the far tail ($y_c \rightarrow 0$) and $\mathcal{E} \simeq 10.8$ for the peak ($y_c \rightarrow 1$) amplitudes. Our final expression for the power spectrum of super-horizon tensor modes is thus
\begin{equation}
   \mathcal{P}_{h,\text{const}}^{\rm DW}(k) = 0.386 \, \mathcal{E}(k \eta_c) \cdot \frac{\gamma^2}{32\pi^2} (f^i_{\rm DW})^2 (\xi k\eta_i)^3 \left( \frac{g_\star}{g_{\star,i}} \right)^{-1/3} \left(\frac{\eta_c}{\eta_i} \right)^7 \,.
\end{equation}
Note that the $k$ dependence of the enhancement factor causes a deviation from perfect $k^3$ causal behavior, which is recovered only in the asymptotic limit $k \rightarrow 0$.

\end{document}